\documentclass[traditabstract]{aa} 
\usepackage{graphicx}
\usepackage{natbib}
\usepackage{longtable}
\newcommand{\HI}{H\,{\sc i}}
\newcommand{\Ha}{H$\alpha$}
\newcommand{\skms}{\ensuremath{\,\mbox{km}\,\mbox{s}^{-1}}}
\newcommand{\kms}{\ensuremath{\mbox{km}\,\mbox{s}^{-1}}}
\usepackage{txfonts}
\begin{document}
   \title{Lopsidedness in WHISP galaxies }

   \subtitle{I. Rotation curves and kinematic lopsidedness}

\titlerunning{Lopsidedness in WHISP galaxies: I. Rotation curves and kinematic lopsidedness}

   \author{J. van Eymeren
          \inst{1}
          \and
          E. J\"utte\inst{2}
	  \and
	  C.~J. Jog\inst{3}
	  \and
	   Y. Stein\inst{2}
	  \and
	   R.-J. Dettmar\inst{2}
          }

   \institute{Fakult\"at f\"ur Physik, Universit\"at Duisburg-Essen, Lotharstr. 1, 47048 Duisburg, Germany\\ 
\email{janine.vaneymeren@uni-due.de}
         \and
            Astronomisches Institut der Ruhr-Universit\"at Bochum, Universit\"atsstr. 150, 44780 Bochum, Germany \\
             \email{eva.juette@astro.rub.de}
\and
Department of Physics, Indian Institute of Science, Bangalore 560012, India
	      \email{cjjog@physics.iisc.ernet.in}
             }

   \date{Accepted 19 March 2011}

 
\abstract{The frequently observed lopsidedness of the distribution of stars
  and gas in disc galaxies is still considered as a major problem in galaxy
  dynamics. It is even discussed as an imprint of the formation history of
  discs and the evolution of baryons in dark matter haloes. Here, we analyse a
  selected sample of 70 galaxies from the Westerbork \HI\ Survey of Spiral and
  Irregular Galaxies. The \HI\ data allow us to follow the morphology and the
  kinematics out to very large radii. In the present paper, we present the
  rotation curves and study the kinematic asymmetry. We extract the rotation
  curves of receding and approaching sides separately and show that the
  kinematic behaviour of disc galaxies can be classified by five different
  types: symmetric velocity fields where the rotation curves of receding and
  approaching sides are almost identical; global distortions where the
  rotation velocities of receding and approaching side have an offset which is
  constant with radius; local distortions which lead to large deviations in
  the inner and negligible deviations in the outer parts (and vice versa); and
  distortions which split the galaxies into two kinematic systems, visible in
  the different behaviour of the rotation curves of receding and approaching
  sides, which leads to a crossing and a change in side. The kinematic
  lopsidedness is measured from the maximum rotation velocities, averaged over
  the plateau of the rotation curves. This gives a good estimate of global
  lopsidedness in the outer parts of the sample galaxies. We find that the
  mean value of the perturbation parameter denoting the lopsided potential as
  obtained from the kinematic data is 0.056. 36\% of all sample galaxies are
  globally lopsided, which can be interpreted as the disc responding to a halo
  that was distorted by a tidal encounter. In Paper II, we study the
  morphological lopsidedness for the same sample of galaxies.}


   \maketitle
%

\section{Introduction}
It has been known for many years that the discs of galaxies often show a large-scale asymmetry in their morphology. Prominent examples are M\,101, NGC\,891 and NGC\,4565 . This asymmetry was for the first time described by \citet{Baldwin1980} and was defined as 'lopsidedness' by these authors. The first systematic study was done 15 years later by \citet{Rix1995} who investigated near-infrared images of a sample of disc galaxies and characterised lopsidedness in the stellar distribution by the $m=1$ mode of a Fourier analysis. They found that at least 30\% of the stellar discs of galaxies are significantly lopsided. A comparable study of gaseous discs was difficult to obtain since spatially resolved maps of a large sample were not available. Therefore, \citet{Richter1994} investigated a large number of integrated \HI\ spectra and found that up to 50\% show an asymmetric global \HI\ profile interpreted by lopsidedness. The advantage of this method is that it is relatively easy to analyse a large dataset. However, it is not clear whether an asymmetric \HI\ profile is caused by an asymmetric gas distribution or by kinematic lopsidedness. The first two-dimensional Fourier decomposition of \HI\ surface density maps, analogous to the stellar studies \citep[e.g., by][]{Rix1995}, was done by \citet{Angiras2006,Angiras2007}, who showed that the HI gas distribution in group galaxies is highly lopsided. For more details on lopsided galaxies see the review by \citet{Jog2009}.

Lopsidedness is observed not only in the morphology of galaxies, but also in the kinematics. Rotation curves are usually derived to obtain the mass distribution in a galaxy that is rotationally supported \citep[e.g.,][]{Binney1987}. The typical assumption is that the rotation curves are azimuthally symmetric. However, observations show local deviations from a smooth circular rotation of a few \kms, which are contributed to streaming motions caused by spiral arms \citep{Shane1966} or bars \citep{Rhee2004}. In addition, it was shown by \citet{Huchtmeier1975} that the rotation curves derived from the two halves of a galactic disc are also asymmetric on large scales. The difference in the rotation velocities was found to be $\geq$ 20\skms. Most recent studies reveal that large-scale kinematic asymmetries are a common phenomenon, as shown by \HI\ studies \citep[e.g.,][]{Kannappan2001,Swaters2009} and also by \Ha\ studies \citep[e.g., the GHASP work by][]{Epinat2008}.

The origin of the asymmetry in the rotation curves has also been addressed theoretically \citep{Jog2002}, where it is shown that the global asymmetry in the rotation curves as well as in the morphology can be caused by the stars and the gas in a galactic disc responding to an imposed distorted halo potential. Even a small lopsided perturbation potential results in highly disturbed kinematics \citep{Jog1997,Jog2002} which should therefore be easy to detect.

An estimate of the perturbation in a lopsided potential that gives rise to the kinematic lopsidedness can be retrieved from the rotation curves of receding and approaching sides. Therefore, it is crucial to accurately define the kinematic parameters, i.e., the systemic velocity, the coordinates of the dynamic centre, the inclination and the position angle. As the perturbation parameter for the lopsided potential is calculated from the maximum rotation velocities, sensitive data which trace the gas up to large radii, where the rotation curve ends in a plateau, are needed. Interferometric \HI\ observations are best suited for this kind of study.

In this first of two papers we introduce the sample, derive rotation curves by
performing a tilted-ring analysis and try to quantify the kinematic
lopsidedness. The morphological lopsidedness will be analysed in a second,
companion paper \citep[][Paper\,II from now on]{vanEymeren2011}.\\

This paper is organised as follows: in Sect.~2 the data selection process is explained, in Sect.~3 we describe the data analysis. In Sect.~4 we present the results, which is followed by a discussion in Sect.~5 and a brief summary in Sect.~6.


\section{Sample selection}
\label{Sectsample}
This paper is based on data from the Westerbork \HI\ Survey of Spiral and Irregular Galaxies (WHISP\footnote{http://www.astro.rug.nl/$\sim$whisp/}). A detailed description of the survey and the data reduction process is given in \citet{Swaters2002}. Three cubes of different spatial resolution per galaxy were produced by the WHISP team, a full resolution cube, one cube smoothed to 30\arcsec\,$\times$\,30\arcsec\ and one cube smoothed to 60\arcsec\,$\times$\,60\arcsec. In order to combine reasonably high spatial resolution with sufficiently high signal to noise, the following analysis is based on the 30\arcsec\,$\times$\,30\arcsec\ data.

We only included galaxies with inclinations ranging between 20\degr\ and 75\degr. This range has been chosen for the following reasons: the velocity information extracted from a nearly face-on galaxy is reduced, which results in a larger uncertainty in measuring the inclination. On the other hand, a nearly edge-on galaxy will provide reliable velocity information, but the surface density map will not be suitable for the analysis of the morphological lopsidedness. Therefore, the above mentioned range of inclinations seemed to be a good compromise. 

In order to be able to measure lopsidedness out to large radii, we further limited the sample to galaxies with the ratio of the \HI\ diameter over the beam size being larger than 10, using the 30\arcsec\,$\times$\,30\arcsec\ data. From the resulting sample we had to eliminate about 30 galaxies either because the signal to noise ratio was low causing the \HI\ intensity distribution to be very patchy or because the velocity field was too distorted to allow for an analysis on global scales. This also excluded galaxies with pronounced warps, visible in the morphology of the nearly edge-on cases or in the velocity field of galaxies of low inclination. This left us with 70 galaxies. Some general properties of the final sample galaxies are given in Table~\ref{Genparm}.

\begin{figure}
 \centering
\includegraphics[width=.35\textwidth,viewport=68 180 260 703, clip=]{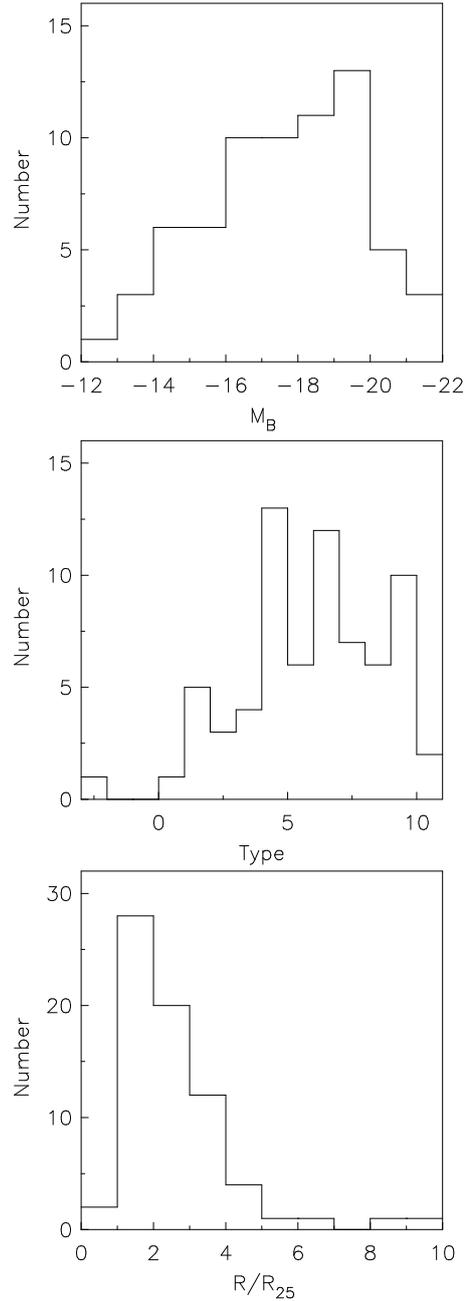}
\caption[]{The distribution of the selected galaxies over absolute \emph{B}-magnitude (upper panel), morphological type (middle panel) and radial extent (lower panel).}
\label{fig_distribution}
\end{figure}

In order to make sure that these selection criteria do not bias our sample towards bright objects and only certain morphological types, we had a look at the range of absolute \emph{B}-magnitudes and morphological types. We found that the galaxies are evenly distributed over -23 to -15 \emph{B}-magnitudes (Fig.~\ref{fig_distribution}, upper panel). We cover a whole range of morphological types with a few early-type spirals and with an increasing number density towards late-type spiral and irregular galaxies (Fig.~\ref{fig_distribution}, middle panel). Most galaxies show emission which is extended out to 1 to 4\,$R_{25}$ (Fig.~\ref{fig_distribution}, lower panel). However, our sample also covers a few galaxies whose \HI\ emission could be detected out to 6 to 10\,$R_{25}$ ($R_{25}$ being the apparent optical radius).
%
\begin{table*}
\caption{\label{Genparm}General parameters of all sample galaxies.}
$$
\begin{tabular}{p{1.5cm}p{1.5cm}p{1.7cm}p{1.5cm}p{1.7cm}p{1.5cm}p{1.5cm}p{1.5cm}p{1.5cm}p{1.5cm}p{1.5cm}}
\hline\hline
\noalign{\smallskip}
Galaxy	& Other name & Hubble Type & $\alpha$ (J2000.0) & $\delta$ (J2000.0) & $M_B$ & $D$ & $R_{25}$ & $R_{\rm max}$\\
& & & [h m s] & [\degr\ \arcmin\ \arcsec] & [mag] & [Mpc] & [kpc] & [$R/R_{25}$]\\
(1) & (2) & (3) & (4) & (5) & (6) & (7) & (8) & (9)\\
\hline
\noalign{\smallskip}
UGC\,625	& IC\,65 & 4 & 01 00 55.4 & +47 40 55.1 & -19.20 & 37.3$^H$ & 13.94 & 2.33\\
UGC\,731	& ... & 9.9 & 01 10 44.0 & +49 36 07.9 & -12.74 & 8.0$^S$ & 2.17 & 3.49\\
UGC\,1249	& IC\,1727 & 8.9 & 01 47 29.9 & +27 20 00.1 & -16.33 & 7.5$^S$ & 7.05 & 1.01\\
UGC\,1256	& NGC\,672 & 6 & 01 47 54.5 & +27 25 58.0 & -17.37 & 7.2$^E$ & 7.42 & 1.20 \\
UGC\,1281	& ... & 7.5 & 01 49 32.0 & +32 35 23.0 & -13.81 & 5.5$^S$ & 4.10 & 1.27 \\
UGC\,1317	& NGC\,674 & 4.9 & 01 51 17.6 & +22 21 28.7 & -20.34 & 42.2$^E$ & 23.33 & 1.84\\
UGC\,1501	& NGC\,784 & 7.8 & 02 01 16.9 & +28 50 14.1 & -14.97 & 5.7$^H$ & 3.46 & 1.80 \\
UGC\,1913	& NGC\,925 & 7 & 02 27 16.9 & +33 34 45.0 & -18.74 & 9.3$^E$ & 14.52 & 1.07 \\
UGC\,2034	& ... & 9.8 & 02 33 43.0 & +40 31 41.2 & -15.82 & 10.1$^S$ & 4.43 & 1.99 \\
UGC\,2080	& IC\,239 & 6 & 02 36 27.9 & +38 58 11.7 & -18.69 & 13.7$^E$ & 8.50 & 2.70 \\
UGC\,2455	& NGC\,1156 & 9.9 & 02 59 42.2 & +25 14 14.2 & -16.26 & 7.8$^S$ & 3.27 & 2.08\\
UGC\,2800	& ... & 10 & 03 40 02.5 & +71 24 21.1 & -14.57 & 20.6$^E$ & 7.02 & 2.56 \\
UGC\,2855	& ... & 5 & 03 48 20.7 & +70 07 58.4 & -17.61 & 17.5$^E$ & 9.03 & 1.55 \\
UGC\,2953	& IC\,356 & 3.4	& 04 07 46.9 & +69 48 44.8 & -19.08 & 15.1$^N$ & 8.74 & 2.39\\
UGC\,3273	& ... & 9 & 05 17 44.4 & +53 33 04.6 & -14.86 & 12.2$^E$ & 4.56 & 2.92 \\
UGC\,3371	& ... &  9.9 & 05 56 38.6 & +75 18 58.0 & -15.57 & 12.8$^S$ & 7.08 & 1.45\\
UGC\,3574	& ... &  5.8 & 06 53 10.4 & +57 10 40.0 & -18.31 & 21.8$^E$ & 4.69 & 3.72 \\
UGC\,3580	& ... & 1.1 & 06 55 30.9 & +69 33 47.0 & -18.40 & 19.2$^N$ & 5.97 & 3.27 \\
UGC\,3734	& NGC\,2344 & 4.4 & 07 12 28.7 & +47 10 00.1 & -17.85 & 15.9$^E$ & 4.83 & 2.63\\
UGC\,3851	& NGC\,2366 & 9.8 & 07 28 54.7 & +69 12 56.8 & -13.80 & 3.4$^S$ & 2.16 & 2.86 \\
UGC\,4173	& ... & 9.9 & 08 07 12.1 & +80 07 30.0 & -15.09 & 16.8$^S$ & 1.54 & 9.50 \\
UGC\,4278	& IC\,2233 & 6.4 & 08 13 58.9 & +45 44 31.7 & -16.31 & 10.5$^S$ & 4.40 & 2.08 \\
UGC\,4284	& NGC\,2541 & 6 & 08 14 40.1 & +49 03 42.2 & -17.15 & 9.8$^E$ & 4.30 & 3.48 \\
UGC\,4458	& NGC\,2599 & 1 & 08 32 11.3 & +22 33 38.0 & -21.01 & 64.2$^N$ & 14.46 & 3.55\\
UGC\,4543	& ... & 7.9 & 08 43 21.6 & +45 44 08.4 & -17.87 & 30.3$^S$ & 5.55 & 4.37\\
UGC\,4838	& NGC\,2776 & 5.2 & 09 12 14.5 & +44 57 17.4 & -20.67 & 37.3$^H$ & 11.59 & 2.57 \\
UGC\,5079	& NGC\,2903 & 4 & 09 32 10.1 & +21 30 03.0 & -19.76 & 8.9$^H$ & 15.25 &  1.40\\
UGC\,5251	& NGC\,3003 & 4.3 & 09 48 36.1 & +33 25 17.4 & -19.23 & 21.5$^E$ & 14.97 & 1.36 \\
UGC\,5253	& NGC\,2985 & 2.4 & 09 50 22.2 & +72 16 43.1 & -20.14 & 21.1$^E$ & 11.14 & 1.65\\
UGC\,5532	& NGC\,3147 & 3.9 & 10 16 53.7 & +73 24 02.7 & -21.53 & 41.1$^E$ & 24.35 & 1.35 \\
UGC\,5685	& NGC\,3254 & 4 & 10 29 19.9 & +29 29 30.6 & -18.96 & 19.9$^H$ & 6.78 & 2.77 \\
UGC\,5717	& NGC\,3259 & 3.7 & 10 32 34.9 & +65 02 27.9 & ... & 25.7$^H$ & 6.49 & 4.60 \\
UGC\,5721	& NGC\,3274 & 6.6 & 10 32 17.3 & +27 40 07.6 & -16.15 & 6.7$^S$ & 1.58 & 4.62 \\
UGC\,5789	& NGC\,3319 & 6 & 10 39 09.5 & +41 41 12.0 & -18.58 & 14.1$^E$ & 7.44 & 2.20 \\
UGC\,5829	& ... & 9.8 & 10 42 41.9 & +34 26 56.0 & -15.98 & 9.0$^S$ & 5.85 & 1.46 \\
UGC\,5918	& ... & 10 & 10 49 36.5 & +65 31 50.0 & -13.27 & 7.7$^S$ & 2.81 & 2.19 \\
UGC\,5997	& NGC\,3403 & 4  & 10 53 54.9 & +73 41 25.3 & -18.25 & 20.4$^H$ & 8.17 & 2.54 \\
UGC\,6225	& NGC\,3556 & 6 & 11 11 31.0 & +55 40 26.8 &  -19.25 & 12.4$^H$ & 7.18 & 2.26 \\
UGC\,6446	& ... & 6.6 & 11 26 40.5 & +53 44 48.0 & -16.22 & 12.0$^S$ & 2.46 & 3.89 \\
UGC\,6537	& NGC\,3726 & 5.1 & 11 33 21.1 & +47 01 45.1 & -19.49 & 14.3$^E$ & 10.92 & 1.33 \\
UGC\,6787	& NGC\,3898 & 1.7 & 11 49 15.4 & +56 05 03.7 & -19.47 & 18.9$^E$ & 9.53 & 2.88 \\
UGC\,6937	& NGC\,3992 & 4 & 11 57 36.0 & +53 22 28.3 & -20.22 & 17.1$^H$ & 20.24 & 0.86 \\
UGC\,7081	& NGC\,4088 & 4.7 & 12 05 34.2 & +50 32 20.5 & -18.96 & 13.2$^H$ & 13.60 & 0.99 \\
UGC\,7090	& NGC\,4095 & 5.3 & 12 06 01.1 & +47 28 42.4 & -18.00 & 10.6$^H$ & 8.67 & 1.07 \\
UGC\,7095	& NGC\,4100 & 4.1 & 12 06 08.5 & +49 34 57.7 & -18.98 & 17.4$^H$ & 11.57 & 1.42 \\
UGC\,7151	& NGC\,4144 & 6 & 12 09 58.6 & +46 27 25.8 & -15.78 & 3.5$^S$ & 2.67 & 1.14 \\
UGC\,7256	& NGC\,4203 & -2.7 & 12 15 05.1 & +33 11 50.4 & -19.09 & 16.9$^N$ & 8.33 & 2.66 \\
UGC\,7321	& ... & 6.6 & 12 17 34.0 & +22 32 24.5 & -14.62 & 7.2$^H$ & 5.01 & 1.25 \\
UGC\,7323	& NGC\,4242 & 7.9 & 12 17 30.2 & +45 37 09.5 & -17.86 & 8.1$^S$ & 4.48 & 1.58 \\
UGC\,7353	& NGC\,4258 & 4 & 12 18 57.5 & +47 18 14.3 & -19.87 & 7.8$^H$ & 19.46 & 1.27 \\
UGC\,7524	& NGC\,4395 & 8.9 & 12 25 48.9 & +33 32 48.7 & -17.59 & 3.5$^S$ & 2.12 & 3.84 \\
UGC\,7603	& NGC\,4455 & 7 & 12 28 44.1 & +22 49 13.6 & -16.80 & 6.8$^S$ & 1.60 & 4.32 \\
UGC\,7766	& NGC\,4559 & 6	& 12 35 57.7 & +27 57 35.1 & -19.81 & 13.0$^E$ & 19.84 & 1.58 \\
UGC\,7989	& NGC\,4725 & 2.2 & 12 50 26.6 & +25 30 02.7 & -21.01 & 18.2$^N$ & 25.91 & 1.07 \\
UGC\,8863	& NGC\,5377 & 1 & 13 56 16.7 & +47 14 08.5 & -19.74 & 27.2$^E$ & 14.36 & 1.65 \\
UGC\,9133	& NGC\,5533 & 2.4 & 14 16 07.7 & +35 20 37.8 & -20.95 &
54.3$^N$ & 22.77 & 3.47 \\
UGC\,9211	& ... & 9.9 & 14 22 32.2 & +45 23 01.9 & -14.89 & 12.6$^S$ & 1.16 & 8.72 \\
UGC\,9649	& NGC\,5832 & 3 & 14 57 45.7 & +71 40 56.4 & -16.06 & 7.7$^E$
& 2.08 & 3.76 \\
UGC\,9858	& ... & 4 & 15 26 41.5 & +40 33 52.2 & ... & 38.2$^E$ & 21.61 & 1.29 \\
\end{tabular}
$$
\end{table*}
\begin{table*}
$$
\begin{tabular}{p{1.5cm}p{1.5cm}p{1.7cm}p{1.5cm}p{1.7cm}p{1.5cm}p{1.5cm}p{1.5cm}p{1.5cm}p{1.5cm}p{1.5cm}}
UGC\,10359	& NGC\,6140 & 5.6 & 16 20 58.2 & +65 23 26.0 & -18.70 & 16.0$^E$ & 4.86 & 3.83\\
UGC\,10470	& NGC\,6217 & 4 & 16 32 39.2 & +78 11 53.4 & -19.59 & 21.2$^E$ & 6.90 & 3.35 \\
UGC\,11670	& NGC\,7013 & 0.5 & 21 03 33.6 & +29 53 50.6 & -17.77 & 12.7$^E$ & 7.70 & 1.56 \\
UGC\,11707	& ... & 8 & 21 14 31.8 & +26 44 04.5 & -16.00 & 15.9$^S$ & 2.37 & 6.35 \\
UGC\,11852	& ... & 1 & 21 55 59.3 & +27 53 54.3 & -19.83 & 80.0$^E$ & 10.61 & 5.48 \\
UGC\,11861	& ... & 7.8 & 21 56 19.4 & +73 15 13.6 & -17.37 & 25.1$^S$ & 6.49 & 3.09 \\
UGC\,11891	& ... & 9.9 & 22 03 33.9 & +43 44 57.2 & -14.79 & 9.0$^E$ & 4.33 & 2.26 \\
UGC\,12082	& ... & 8.8 & 22 34 10.8 & +32 51 37.8 & -16.13 & 10.1$^E$ & 3.95 & 2.04 \\
UGC\,12632	& ... & 8.7  & 23 29 58.7 & +40 59 24.8 & -16.06 & 6.9$^S$ & 4.28 & 1.88 \\
UGC\,12732	& ... & 8.7 & 23 40 39.9 & +26 14 11.1 & -16.29 & 13.2$^S$ & 5.29 & 2.90 \\
UGC\,12754	& NGC\,7741 & 6 & 23 43 54.4 & +26 04 32.2 & -18.25 & 8.9$^E$ & 4.70 & 1.51 \\
\hline
\end{tabular}
$$	
\footnotesize{ 
Notes: (1) galaxy name from the UGC catalogue; (2) other common names; (3) morphological type following the classification by \citet{deVaucouleurs1979}; (4) and (5) equatorial coordinates of the optical centre (NED); (6) absolute \emph{B}-magnitudes (HyperLeda); (7) distance \emph{D} (H: deduced from the systemic velocity taken from HyperLeda corrected for Virgocentric infall and assuming $H_0=75$ \skms\,Mpc$^{-1}$; S: \citet{Swaters2002}; N: \citet{Noordermeer2005}; E: \citet{Epinat2008}); (8) apparent radius (HyperLeda); (9) outer \HI\ radius as defined in this paper.}
\end{table*}
%
\section{Data analysis}
The data analysis is based on routines within the Groningen Image Processing System \citep[GIPSY\footnote{URL: http://www.astro.rug.nl/$\sim$gipsy/},][]{vanderHulst1992}.
\subsection{Tilted-ring analysis}
The calculation of the kinematic lopsidedness requires the knowledge of the
kinematic parameters of each galaxy. Therefore, a tilted-ring analysis of the
velocity fields created from the peak value of the \HI\ line profiles was
performed (GIPSY task \emph{rotcur}). The following steps were carried out:
initial estimates for the centre coordinates $x_0$ and $y_0$, the inclination
$i$, and the position angle $PA$ were obtained by fitting ellipses to the \HI\
intensity distribution out to 3$\sigma$ (GIPSY task \emph{ellfit}). The
systemic velocity $v_{\rm sys}$ was taken from NED. These values were then fed
into \emph{rotcur} and iteratively processed, always for a combination of
receding and approaching side \citep[see, e.g., ][]{vanEymeren2009}. The width
of the rings was chosen to be half the spatial resolution, i.e.,
15\arcsec. Note that all pixels along a ring have equal weights. The expansion velocity $v_{\rm exp}$ was fixed to zero. We fixed all parameters except for the systemic velocity that was allowed to vary with radius. After running \emph{rotcur}, a mean value of $v_{\rm sys}$ was calculated and fed into \emph{rotcur} as a now best-fitting and fixed value. In this way, we calculated best-fitting values for $v_{\rm sys}$, $x_0$ and $y_0$, $i$, and $PA$ (in the here given order). In a final run, we used the best-fitting values for all parameters, kept them fixed for all radii, and derived the rotation velocities for a combination of receding and approaching side.

As the velocity fields of many galaxies look quite asymmetric, we also derived rotation curves for both sides separately. The systemic velocity $v_{\rm sys}$ and the centre coordinates $x_0$ and $y_0$ were fixed to the values derived in the first run for a combination of receding and approaching side. Inclination and position angle were calculated for each side separately and iteratively, as described above.

As a last step, we checked how well the derived parameters describe the observed velocity fields. Therefore, we created model velocity fields by using the parameters from the final \emph{rotcur} run. Subsequently, the model was subtracted from the original velocity field. Typically, the residuals are of the order of a few \kms, which means that the derived parameters describe the kinematics of the observed galaxies quite well. However, some of the nearly face-on galaxies show very disturbed velocities indicating a warp, which cannot be described by a set of parameters that is constant with radius. On the other side, we also found some almost edge-on galaxies, which show signs of warps in their outer morphology, whereas the velocity field is not affected at all. As already mentioned in Sect.~\ref{Sectsample}, all these galaxies have been removed from the final sample.

Warps in highly-inclined galaxies can easily be detected by looking at the change of scale-height with radius \citep[][]{Schwarzkopf2001,Garcia-Ruiz2002}. In galaxies with low inclination, lopsidedness would dominate the morphological shape so that we can be sure that the contribution of a potential warp is small along the line of sight unless the velocity field is globally twisted as mentioned above. More problematic are the galaxies with intermediate inclination. Here, some contamination from warps while measuring lopsidedness is quite likely.

The large-scale structure of warps has been classified as U-, S-, N-, and L-shaped \citep[e.g.,][]{Reshetnikov1998,Garcia-Ruiz2002}. The first three still cause a symmetric distribution of the gas (or the stars), whereas the latter results in an asymmetric distribution, i.e., the kind of distribution we want to study. This means that for L-shaped warps it is difficult to clearly separate warpiness from lopsidedness (at intermediate inclination). However, optical studies by \citet{Schwarzkopf2001} or \citet{Sanchez-Saavedra2003} and \HI\ studies by \citet{Garcia-Ruiz2002} show that the fraction of galaxies harbouring L-shaped warps is small in comparison to galaxies showing U- or S-shaped warps or no warp at all.
\subsection{Kinematic lopsidedness}
\label{Sectkinlop}
Under the assumption that lopsidedness occurs as a disc response to a distorted halo, which was caused by a galaxy interaction, an estimate of the lopsided perturbation potential that gives rise to the kinematic lopsidedness can be retrieved from the maximum rotation velocities of the sample galaxies \citep{Jog2002}:
\begin{equation}
\epsilon_{\rm kin}=\frac{v_{\rm rec}-v_{\rm appr}}{2v_{\rm c}}.
\label{eqv}
\end{equation}
where $v_{\rm rec}$ and $v_{\rm appr}$ are the maximum rotation velocities, as measured from the plateau of the rotation curves of the receding and approaching sides respectively, $v_{\rm c}$ is the maximum rotation velocity as measured from the plateau of the combined rotation curve, and $\epsilon_{\rm kin}$ is the perturbation parameter that denotes the lopsided potential.

In many galaxies the rotation curves are flat in the outer parts. However, in some galaxies the rotation curves are still rising at the outermost points. In those cases we took the highest value of $v_{\rm c}$, which resulted in an upper limit for $\epsilon_{\rm kin}$.
\section{Results}
\subsection{Rotation curves -- an overview}
\label{Sectrotcur}
\begin{figure*}
 \centering
\includegraphics[width=.68\textwidth,viewport=58 120 390 738, clip=]{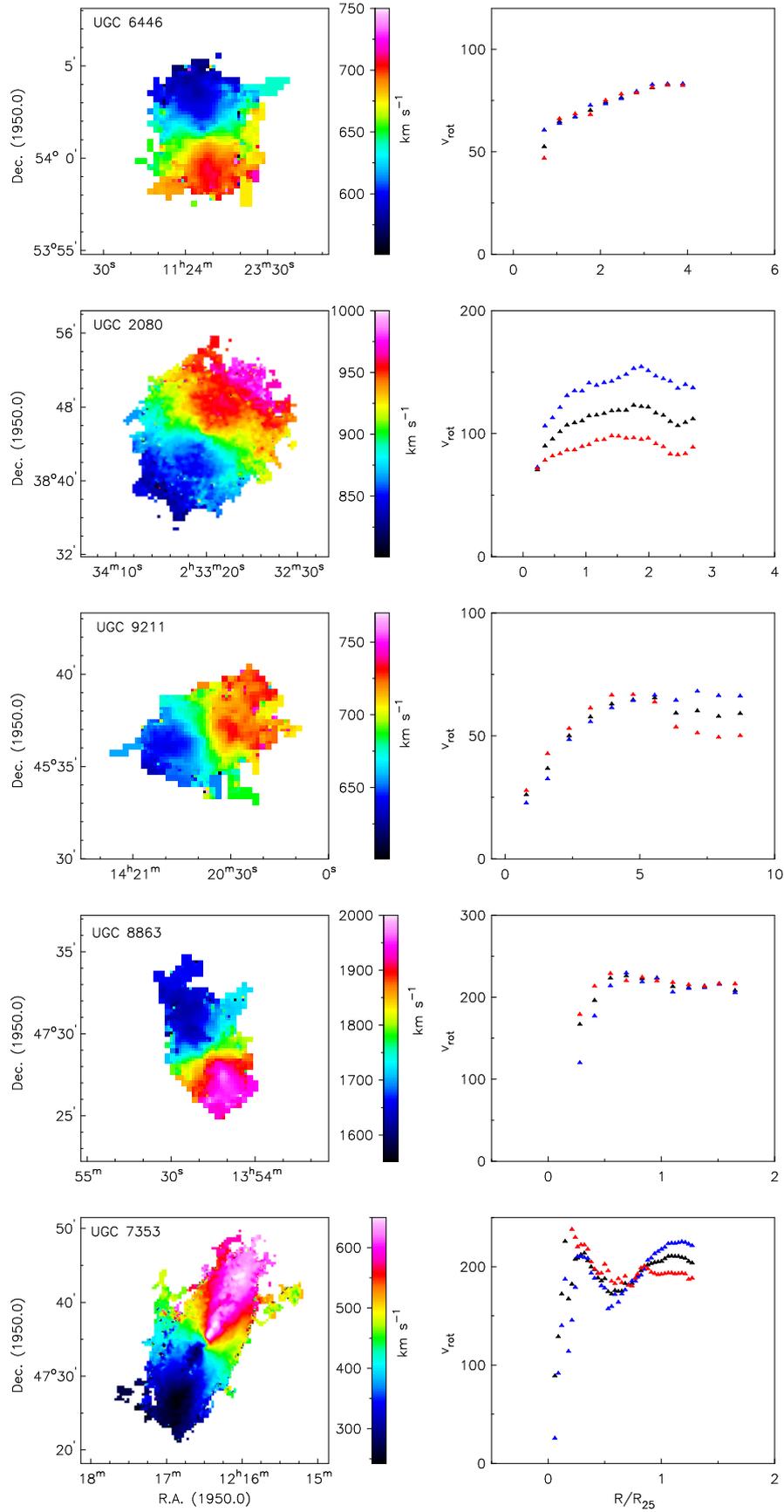}
\caption[]{Velocity fields and rotation curves of some example galaxies representing five different types of velocity patterns. From top to bottom: UGC\,6446 (Type~1: receding and approaching sides agree at all radii), UGC\,2080 (Type~2: receding and approaching sides have a constant offset), UGC\,9211 (Type~3: receding and approaching sides agree well at small radii, but differ significantly at large radii), UGC\,8863 (Type~4: receding and approaching sides differ at small radii, but agree well at large radii), UGC\,7353 (Type~5: the curves of receding and approaching sides change sides).}
\label{fig_examples}
\end{figure*}

Figures~\ref{fig_rotcur1} to \ref{fig_rotcur3} show the rotation curves of all sample galaxies. Black symbols represent the rotation velocities derived by combining receding and approaching sides, blue and red symbols represent the rotation velocities of approaching and receding sides respectively. \citet{Swaters1999} have already shown a type of velocity field in spiral galaxies where receding and approaching sides are distinctly different, e.g., one side keeps rising while the other side declines (see their Fig.~2). This case is here catalogued as Type~5. Beyond this, we have found four additional types of rotation curves, which are discussed below. We give examples for each type in Fig.~\ref{fig_examples}.

\emph{Type~1} For 13 galaxies (about 19\%\ of the whole sample), receding and approaching sides are in very good agreement indicating a very symmetric velocity distribution. Fig.~\ref{fig_examples}, upper row shows as an example the late-type spiral galaxy UGC\,6446. Its velocity field (left panel) is very smooth and shows the typical spider pattern. The derived rotation velocities for receding and approaching sides (right panel) only differ marginally.

\emph{Type~2} For 23 galaxies (about 33\%\ of the whole sample), receding and approaching side have a constant offset. As an example we show the irregular dwarf galaxy UGC\,2080 (Fig.~\ref{fig_examples}, second row). There are no local asymmetries in the velocity field, but a global offset of the iso-velocity contours.

\emph{Type~3} In 10 galaxies (i.e., in 14\%\ of all cases), receding and approaching sides are in good agreement at small radii, but show increasing differences at larger radii. A good example is the irregular dwarf galaxy UGC\,9211 (Fig.~\ref{fig_examples}, third row). Whereas the rotation velocities extracted from the approaching side form a typical rotation curve with a slow rise in the inner parts and a plateau in the outer parts, the rotation velocities extracted from the receding side start to decline significantly from a radius of 6\,$R_{25}$ on. The velocity field shows gas which seems to be attached to the north-western part of the galaxy and probably causes the unexpected shape of the rotation curve of the receding side.

\emph{Type~4} We detected seven cases (which corresponds to 10\%) where receding and approaching sides agree well in the outer parts, but differ significantly in the inner parts. A good example is the barred spiral galaxy UGC\,8863 (Fig.~\ref{fig_examples}, fourth row). The velocity field shows a bar-like feature in the inner parts, which probably leads to these large discrepancies between receding and approaching sides.

\emph{Type~5} Last but not least we found 17 galaxies (corresponding to 24\%) with differences between receding and approaching sides both in the inner and the outer parts. However, the curves change in that the rotation velocities of both sides "meet" and the decreasing side keeps decreasing, whereas the increasing side keeps increasing, which means that the sides flip. As an example we show the rotation curves of the spiral galaxy UGC\,7353 (Fig.~\ref{fig_examples}, lower row). At a radius of about 0.8\,$R_{25}$ the rotation velocities extracted from the approaching side strongly increase, while the rotation velocities extracted from the receding side stay almost constant. The velocity field reveals two systems, an inner one which rotates more quickly in comparison to the outer one. This probably causes the change in the behaviour of the rotation velocities of both sides.
\subsection{Rotation curves -- notes on individual galaxies}
We now want to discuss a few peculiar rotation curves in more detail. We only concentrate on galaxies which are not included in the sample by \citet{Swaters2009}.

\emph{UGC\,1249} shows a constant offset of 50\skms\ between the rotation curves of receding and approaching sides and is therefore classified as Type~2 galaxy. The velocity field reveals a close companion in the north, UGC\,1256, which is at least twice as massive as UGC\,1249. Both galaxies seem to interact with each other. While UGC\,1249 is strongly affected by this process, UGC\,1256 appears to be not lopsided at all (therefore classified as Type~1).

\emph{UGC\,1317} is in a group with four other galaxies (UGC\,1280, UGC\,1286, UGC\,1305, and UGC\,1310). It shows a declining rotation curve in the outer parts. Receding and approaching sides are offset by about 50\skms\ at all radii, which classifies this galaxy as Type~2.

\emph{UGC\,1501} is very symmetric in the inner parts, but shows distortions in the outer parts (Type~3). The velocity field reveals gas which seems to be attached to the western edge. Therefore, gas accretion might be the possible cause of the lopsidedness.

\emph{UGC\,2855} has a close companion, UGC\,2866, which is three times less massive. UGC\,2855 shows a tail in the north-western part and additional gas attached to the southern edge. Receding and approaching sides are in good agreement in the inner parts. However, in the outer parts, the approaching rotation curve shows a decline, whereas the receding one slowly increases (Type~3). The difference in velocity is more than 50\skms. 

\emph{UGC\,4173} is very slowly rotating with only 36\skms\ (there is only one more galaxy, UGC\,2034, in our sample which shows an even lower rotation velocity of about 24\skms). Receding and approaching sides differ by about 10\skms\ across the whole radial range (Type~2). The most remarkable thing about this galaxy is the extent to which we detect the neutral gas, which is out to almost 10\,$R_{25}$. Other examples are UGC\,9211 (about 9$R_{25}$), UGC\,11707 (about 7$R_{25}$), and UGC\,11852 (about 6$R_{25}$). Typically, the \HI\ disc is twice as extended as the optical disc in a normal spiral galaxy \citep{Briggs1980}. Irregular galaxies can have a more extended \HI\ distribution \citep[e.g.,][]{Huchtmeier1981}. However, only one other case of an \HI\ disc, as extended as UGC\,4173, has been published so far, which is NGC\,3714 \citep{Gentile2007}.

\emph{UGC\,4458} also reveals a close companion, which is quite mass-poor (about six times less massive than UGC\,4458). The offset between receding and approaching sides is visible at all radii (Type~2) and very large with 100 to 150\skms. In fact, the rotation velocities are unusually high with values of about 550\skms\ (approaching side) and 350\skms\ (receding side) at 0.75\,$R_{25}$, decreasing at larger radii to 400 and 300\skms\ respectively. However, note that the inclination is very low. From the velocity field it seems that UGC\,4458 interacts with its small companion.

\emph{UGC\,5251} is classified as Type~5. In both the inner and outer parts, receding and approaching sides are offset by about 50\skms\ with a change of sides at a radius of 1\,$R_{25}$. The velocity field is highly asymmetric.

\emph{UGC\,5532} also shows unusually high rotation velocities with a constant offset of about 50\skms\ between receding and approaching sides (Type~2). No close companion could be detected on the \HI\ image. This galaxy has also been observed by the \emph{GHASP}\footnote{Gassendi HAlpha survey of SPirals} team who derived a similarly looking rotation curve from the \Ha\ velocity field \citep{Epinat2008}.

\emph{UGC\,7256} is an extreme case of Type~5 as the rotation velocities of receding and approaching sides change sides several times. The velocity field of UGC\,7256 looks slightly warped. Additionally, we located a small \HI\ cloud nearby.

\emph{UGC\,7353} is also classified as Type~5 as receding and approaching sides change sides at a radius of about 0.8\,$R_{25}$. The offset in the outer part is about 50\skms, it varies in the inner parts, but is generally smaller. \HI\ gas was detected at the eastern edge, a very small companion is located to the north-west.

\emph{UGC\,10470} shows a second rise of the rotation curve at 2.5\,$R_{25}$. The rotation curves of receding and approaching sides are offset by about 40\skms\ at all radii (Type~2).

The combined rotation curve of \emph{UGC\,12082} lies above the rotation curve for receding and approaching side separately. This is due to the very low inclination. 

\emph{UGC\,12754} reveals a declining rotation curve at large radii. The rotation velocities of receding and approaching sides differ in the inner parts and are mainly comparable in the outer parts (Type~4).

\subsection{Kinematic lopsidedness}
As we could show in the previous section, the rotation curves of galaxies allow us to examine the kinematic lopsidedness. The rotation curves of most galaxies end in a plateau so that it is possible to derive a perturbation parameter in the lopsided potential that is measured from the kinematic data, $\epsilon_{\rm kin}$, from the maximum rotation velocities (see Sect.~\ref{Sectkinlop}). As already mentioned in Sect.~\ref{Sectkinlop}, $\epsilon_{\rm kin}$ only gives us an estimate as the maximum rotation velocities are not always well defined. In Table~\ref{Kinpar}, last column, we list $\epsilon_{\rm kin}$ for all sample galaxies.

\begin{figure}
 \centering
\includegraphics[width=.4\textwidth,viewport=68 523 260 703, clip=]{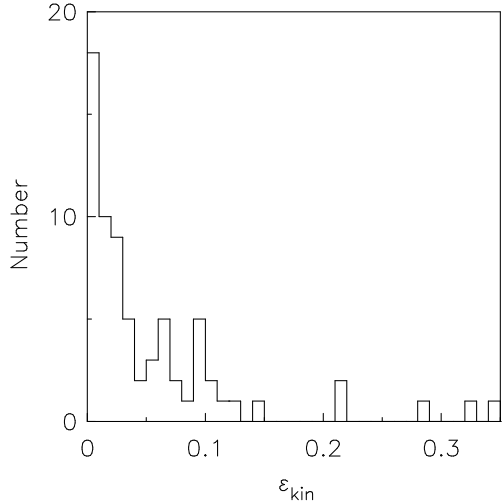}
\caption[]{The distribution of $\epsilon_{\rm kin}$ as calculated from Eq.~\ref{eqv}.}
\label{fig_ev}
\end{figure}
Figure~\ref{fig_ev} shows the distribution of $\epsilon_{\rm kin}$. Although the mean was calculated to be about 0.056, it becomes obvious that most galaxies have values far below the mean, which means that the kinematic lopsidedness is generally quite low. In fact, more than 60\%\ of all sample galaxies show values for the perturbation parameter for the potential $\epsilon_{\rm kin}$ to be below 0.056.
%
\begin{table*}
\caption{\label{Kinpar}Kinematic parameters as obtained from the tilted-ring analysis.}
$$
\begin{tabular}{p{1.2cm}p{1.2cm}p{1.5cm}p{1.7cm}p{1.2cm}p{1.2cm}p{1.2cm}p{1.2cm}p{1.2cm}p{0.8cm}p{1.2cm}}
\hline\hline
\noalign{\smallskip}
UGC & $v_{\rm sys}$ & $\alpha$ (J2000.0) & $\delta$ (J2000.0) & $i$ & $PA$ & $v_{\rm c}$ & $v_{\rm rec}$ & $v_{\rm appr}$ & Type & $\epsilon_{\rm kin}$\\
& [\skms] & [h m s] & [\degr\ \arcmin\ \arcsec] & [\degr] & [\degr] & [\kms] & [\kms] & [\kms] & &\\
(1) & (2) & (3) & (4) & (5) & (6) & (7) & (8) & (9) & (10) & (11)\\
\hline
\noalign{\smallskip}
625 & 2608 & 01 00 55.6 & +47 40 50.8 & 70.27 & 331.74 & 168.52 & 159.48 & 181.41 & 2 & 0.065\\ 
731 & 642 & 01 10 40.4 & +49 36 03.8 & 59.30 & 258.16 & 72.86 & 73.45 & 73.38 & 1 & 0.001\\
1249 & 343 & 01 47 33.7 & +27 20 50.3 & 51.43 & 147.57 & 65.49 & 89.23 & 46.54 & 2 & 0.326\\
1256 & 430 & 01 47 51.3 & +27 25 57.1 & 71.59 & 69.44 & 114.35 & 113.92 & 115.01 & 1 & 0.005\\
1281 & 156 & 01 49 31.9 & +32 35 27.3 & 73.34 & 216.86 & 59.45 & 59.74 & 60.82 & 1 & 0.009\\
1317 & 3101 & 01 51 14.3 & +22 21 44.2 & 71.95 & 107.13 & 229.96 & 226.82 & 230.67 & 2 & 0.008\\
1501 & 196 & 02 01 16.8 & +28 50 05.3 & 73.88 & 0.83 & 53.39 & 49.03 & 58.69 & 3 & 0.091\\
1913 & 556 & 02 27 15.0 & +33 34 55.6 & 53.65 & 286.63 & 114.79 & 117.44 & 114.49 & 4 & 0.013\\
2034 & 577 & 02 33 43.0 & +40 31 45.4 & 29.85 & 358.67 & 24.44 & 13.90 & 30.94 & 2 & 0.349\\
2080 & 908 & 02 36 27.2 & +38 58 15.4 & 27.64 & 335.35 & 117.52 & 145.57 & 94.26 & 2 & 0.218\\
2455 & 372 & 02 59 41.8 & +25 14 29.1 & 52.11 & 271.88 & 51.57 & 47.57 & 58.60 & 2 & 0.107\\
2800 & 1177 & 03 40 08.0 & +71 24 17.9 & 61.87 & 285.49 & 113.26 & 111.75 & 114.19 & 3 & 0.011\\
2855 & 1195 & 03 48 24.8 & +70 08 19.6 & 61.49 & 111.19 & 205.32 & 207.06 & 215.02 & 3 & 0.019\\
2953 & 873 & 04 07 46.8 & +69 48 51.2 & 52.18 & 100.83 & 291.07 & 289.47 & 305.54 & 5 & 0.028\\
3273 & 617 & 05 17 44.9 & +53 33 03.6 & 74.04 & 40.59 & 93.86 & 95.01 & 93.25 & 3 & 0.009\\
3371 & 814 & 05 56 36.4 & +75 19 00.4 & 45.52 & 130.19 & 84.50 & 78.11 & 82.86 & 2 & 0.028\\
3574 & 1441 & 06 53 10.7 & +57 10 35.5 & 31.20 & 99.79 & 125.98 & 108.89 & 123.71 & 3 & 0.059\\
3580 & 1194 & 06 55 30.7 &  +69 33 40.1 & 65.68 & 1.87 & 116.09 & 112.35 & 119.34 & 1 & 0.030\\
3734 & 972 & 07 12 28.5  & +47 10 03.9 & 21.61 & 135.22 & 172.19 & 177.11 & 133.61 & 2 & 0.126\\
3851 & 104 & 07 28 54.6 &  +69 12 47.4 & 67.59 & 38.76 & 51.28 & 54.62 & 49.26 & 5 & 0.052\\
4173 & 859 & 08 07 05.6 &  +80 07 43.7 & 69.39 & 135.05 & 35.61 & 28.24 & 38.90 & 2 & 0.150\\
4278 & 556 & 08 13 59.0 &  +45 44 31.0 & 64.17 & 352.32 & 87.50 & 88.50 & 87.35 & 1 & 0.007\\
4284 & 557 & 08 14 40.5 &  +49 03 41.0 & 61.67 & 169.77 & 104.06 & 103.21 & 104.17 & 5 & 0.005\\
4458 & 4753 & 08 32 11.3 & +22 33 34.0 & 22.10 & 289.29 & 292.78 & 400.03 & 272.95 & 2 & 0.217\\
4543 & 1965 & 08 43 20.3 &  +45 43 59.5 & 61.58 & 316.35 & 53.31 & 68.72 & 38.01 & 2 & 0.288\\
4838 & 2624 & 09 12 16.8 &  +44 57 17.5 & 29.14 & 304.42 & 150.33 & 129.71 & 131.79 & 2 & 0.007\\
5079 & 551 & 09 32 10.3 &  +21 30 10.4 & 64.01 & 202.69  & 195.31 & 193.13 & 198.56 & 4 & 0.014\\
5251 & 1497 & 09 48 34.3 &  +33 25 15.0 & 64.50 & 253.99 & 142.61 & 156.03 & 127.52 & 5 & 0.100\\
5253 & 1324 & 09 50 21.4 &  +72 16 37.8 & 36.32 & 355.95 & 249.47 & 277.60 & 227.78 & 2 & 0.100\\
5532 & 2826 & 10 16 55.7 &  +73 23 57.7 & 35.19 & 142.79 & 300.51 & 284.11 & 260.47 & 2 & 0.039\\
5685 & 1334 & 10 29 19.1 &  +29 29 15.5 & 72.67 & 45.21 & 205.25 & 189.77 & 227.31 & 5 & 0.091\\
5717 & 1687 & 10 32 35.5 &  +65 02 56.7 & 52.30 & 10.76 & 131.30 & 132.40 & 127.69 & 5 & 0.018\\
5721 & 537 & 10 32 17.1 &  +27 40 08.6 & 65.53 & 279.45 & 80.29 & 76.31 & 86.31 & 2 & 0.062\\
5789 & 738 & 10 39 09.6 &  +41 41 12.9 & 62.49 & 35.56 & 105.14 & 104.51 & 107.04 & 1 & 0.012\\
5829 & 628 & 10 42 43.0 &  +34 27 26.0 & 36.94 & 191.50 & 53.79 & 48.46 & 47.94 & 5 & 0.005\\
5918 & 335 & 10 49 36.0 &  +65 31 58.3 & 54.04 & 238.60 & 38.07 & 36.22 & 43.82 & 3 & 0.100\\
5997 & 1266 & 10 53 55.4 &  +73 41 25.9 & 66.43 & 71.60 & 149.75 & 153.97 & 146.11 & 1 & 0.026\\
6225 & 702 & 11 11 31.1 &  +55 40 20.0 & 75.87 & 256.19 & 157.44 & 160.81 & 155.47 & 4 & 0.017\\
6446 & 646 & 11 26 40.7 &  +53 44 51.0 & 50.49 & 191.11 & 81.35 & 82.53 & 81.58 & 1 & 0.006\\
6537 & 860 & 11 33 20.8 &  +47 01 50.8 & 49.47 & 195.65 & 158.52 & 147.23 & 155.79 & 5 & 0.027\\
6787 & 1182 & 11 49 15.5 &  +56 05 01.6 & 68.40 & 113.51& 237.54 & 247.13 &  228.86 & 2 & 0.039\\
6937 & 1056 & 11 57 35.7 &  +53 22 24.5 & 57.18 & 248.17 & 268.79 & 269.50 & 271.25 & 1 & 0.003\\
7081 & 748 & 12 05 28.2 &  +50 32 33.4 & 65.55 & 228.38 & 179.40 & 166.86 & 187.94 & 2 & 0.059\\
7090 & 557 & 12 06 00.6 &  +47 28 28.3 & 67.38 & 16.54 & 156.93 & 139.02 & 176.52 & 5 & 0.120\\
7095 & 1081 & 12 06 08.3 &  +49 35 04.0 & 70.52 & 345.65 & 187.04 & 196.50 & 181.48 & 2 & 0.040\\
7151 & 266 & 12 09 58.7 &  +46 27 24.4 & 74.89 & 281.33 & 83.55 & 85.22 & 79.92 & 1 & 0.032\\
7256 & 1098 & 12 15 05.5 &  +33 11 30.3 & 50.50 & 204.18 & 144.96 & 155.53 & 133.76 & 5 & 0.075\\
7321 & 409 & 12 17 33.7 &  +22 32 25.5 & 74.90 & 260.55 & 106.34 & 106.73 & 104.34 & 3 & 0.011\\
7323 & 509 & 12 17 29.4 &  +45 36 50.5 & 52.06 & 36.02 & 71.95 & 72.64 & 74.05 & 4 & 0.010\\
7353 & 460 & 12 18 56.8 &  +47 18 30.1 & 69.57 & 331.25 & 200.53 & 209.37 & 192.24 & 5 & 0.043\\
7524 & 317 & 12 25 50.4 &  +33 32 35.0 & 46.87 & 324.87 & 76.03 & 73.90 & 74.19 & 5 & 0.002\\
7603 & 650 & 12 28 43.5 &  +22 50 09.1 & 73.12 & 199.93 & 65.07 & 68.97 & 60.59 & 2 & 0.064\\
7766 & 805 & 12 35 58.5 &  +27 57 16.9 & 67.31 & 323.31 & 121.95 & 114.51 & 131.16 & 5 & 0.068\\
7989 & 1190 & 12 50 25.8 &  +25 29 56.2 & 44.19 & 32.45 & 256.47 & 243.12 & 242.60 & 4 & 0.001\\
8863 & 1787 & 13 56 16.3 &  +47 14 07.3 & 54.06 & 211.35 & 213.60 & 211.17 & 215.06 & 4 & 0.009\\
9133 & 3841 & 14 16 06.9 &  +35 20 21.6 & 51.94 & 30.36 & 237.97 & 231.80 &
245.23 & 3 & 0.028\\
9211 & 688 & 14 22 32.0 &  +45 23 05.0 & 46.22 & 291.13 & 61.60 & 63.72 & 60.02 & 3 & 0.030\\
9649 & 446 & 14 57 47.3  & +71 40 54.7 & 56.56 & 228.21 & 88.27 & 89.76 &
89.34 & 1 & 0.002\\
9858 & 2628 & 15 26 41.5 &  +40 33 53.7 & 69.98 & 75.80 & 170.73 & 184.91 & 154.54 & 5 & 0.089\\
\end{tabular}
$$
\end{table*}
\begin{table*}
$$
\begin{tabular}{p{1.2cm}p{1.2cm}p{1.5cm}p{1.7cm}p{1.2cm}p{1.2cm}p{1.2cm}p{1.2cm}p{1.2cm}p{0.8cm}p{1.2cm}}
10359 & 908 & 16 20 57.0 &  +65 23 32.9 & 43.82 & 278.01 & 127.97 & 122.73 & 126.05 & 1 & 0.013\\
10470 & 1359 & 16 32 39.8 &  +78 12 07.9 & 36.78 & 300.11 & 130.96 & 142.00 & 117.28 & 2 & 0.094\\
11670 & 777 & 21 03 33.4 &  +29 53 49.9 & 67.34 & 335.63 & 173.00 & 177.36 & 168.55 & 2 & 0.026\\
11707 & 901 & 21 14 30.8 &  +26 44 04.1 & 65.37 & 55.72 & 97.71 & 89.37 & 102.82 & 5 & 0.069\\
11852 & 5837 & 21 55 59.6 &  +27 53 57.0 & 46.03 & 192.35 & 181.95 & 163.18 & 171.37 & 2 & 0.023\\
11861 & 1481 & 21 56 22.3 &  +73 15 40.6 & 48.30 & 216.95 & 152.18 & 149.95 & 148.84 & 1 & 0.004\\
11891 & 457 & 22 03 31.4 &  +43 45 11.1 & 46.26 & 115.49 & 87.50 & 81.05 & 84.74 & 5 & 0.021\\
12082 & 805 & 22 34 11.3 &  +32 51 24.3 & 20.56 & 145.25 & 87.19 & 58.33 & 45.02 & 2 & 0.076\\
12632 & 426 & 23 30 00.6 &  +40 59 41.5 & 49.49 & 38.38 & 65.93 & 69.07 & 69.00 & 5 & 0.001\\
12732 & 747 & 23 40 40.1 &  +26 14 00.0 & 40.27 & 14.33 & 79.34 & 83.81 & 78.05 & 3 & 0.036\\
12754 & 753 & 23 43 54.1 &  +26 04 38.5 & 49.46 & 344.85 & 116.74 & 116.93 & 120.25 & 4 & 0.014\\
\hline
\end{tabular}
$$	
\footnotesize{ 
Notes: (1) the UGC number; (2) to (6) the kinematic parameters as iteratively calculated from the tilted-ring analysis for a combination of receding and approaching sides: (2) the systemic velocity, (3) and (4) equatorial coordinates of the dynamic centre, (5) the inclination; (6) the position angle which is measured in anti-clockwise direction from the north to the receding half of the galaxy; (7) to (9) the maximum rotation velocities as measured from the plateaus of the rotation curves: $v_{\rm c}$ (for a combination of receding and approaching sides), $v_{\rm rec}$ (as measured from the receding side), and $v_{\rm appr}$ (as measured from the approaching side); (10) the type of rotation curve as defined in Sect.\ref{Sectrotcur}; (11) the perturbation parameter in the lopsided potential as measured from the kinematic data (see Eq.~\ref{eqv}).}
\end{table*}
\section{Discussion}
We analysed the rotation curves of 70 spiral and irregular galaxies. In many cases, the rotation curves extracted from receding and approaching sides separately differ significantly. Galaxies of Type~2 (like UGC\,1249 or UGC\,1317), where receding and approaching sides are offset at all radii, either have a more massive companion or are part of a group of close companions. Type~3 galaxies, i.e., galaxies which are kinematically symmetric in the inner parts, but asymmetric in the outer parts (just from comparing the rotation velocities of receding and approaching sides), often show close companions, which are typically less massive, or small \HI\ clouds. Close companions are also seen for Type~5 galaxies. In this class, the curves of receding and approaching sides show a different behaviour and therefore cross each other and change sides. This indicates that the detailed kinematic asymmetry including the spatial extent of lopsidedness strongly depends on the environment of galaxies. Type~2 and Type~5 galaxies show lopsidedness at all radii. This is expected if the disc responds to a halo that was distorted by a tidal encounter \citet{Jog1997}. As those two types of galaxies form 57\% of the whole sample, we suggest that tidal encounters play an important role for the lopsidedness of disc galaxies. The detection of companions and \HI\ clouds in the near vicinity of the here studied galaxies supports this scenario. In addition, gas accretion as well as ram pressure from the intergalactic medium could also play a role for the origin of lopsidedness, as, e.g., simulated by \citet{Mapelli2008}. From a comparison with NGC\,891 they find that in this particular galaxy tidal interaction seems to be the dominant process for generating lopsidedness.

Our data confirm the results by \citet{Huchtmeier1975} of an offset of receding and approaching sides larger than 20\skms. In fact, our sample galaxies often reveal offsets as large as 50\skms.

We now want to check how well $\epsilon_{\rm kin}$ represents the observed asymmetries in the velocity fields. Therefore, we concentrate on our example galaxies displayed in Fig.~\ref{fig_examples} and compare the differences of the rotation curves of receding and approaching halves with the perturbation parameter in the lopsided potential. In UGC\,6446, the rotation velocities of receding and approaching sides are in very good agreement at all radii. Not surprisingly, the value of $\epsilon_{\rm kin}$ is close to zero. In UGC\,9211, $\epsilon_{\rm kin}$ is clearly below the mean. However, the parameter does not account for the large discrepancies in the very outer parts of the galaxy. For UGC\,2080, $\epsilon_{\rm kin}$ is very high, which reflects very nicely the constant, large difference between receding and approaching side. The inner parts of UGC\,8863 show large differences in the rotation velocities of receding and approaching side. The perturbation parameter in the potential, however, is far below the mean because the velocities were averaged over the plateau where the differences are small. UGC\,7353 does not have a clearly defined plateau. The maximum rotation velocities were taken from the first maximum of the rotation curves where receding and approaching sides differ only slightly. Therefore, the perturbation parameter in the potential is small, although large distortions are visible both in the very inner and very outer parts of the galaxy.

Taking into account the results for all sample galaxies, we conclude that $\epsilon_{\rm kin}$ is a very reliable parameter if a galaxy is globally distorted or not distorted at all. As the maximum rotation velocities are taken from a limited radial range (where the rotation curves reach their maximum), local distortions are either not included or overestimated, depending on whether we have a Type~4 or Type~3 galaxy. In about 20\%\ of all sample galaxies, the \HI\ content does not extend much further out than $R_{25}$ so that the maximum rotation velocities were averaged over a radial range within $R_{25}$. In all other galaxies, the maximum rotation velocities were measured at radii beyond $R_{25}$.

Summarised, it can be said that the best way to quantify the kinematic lopsidedness is a comparison of the rotation curves of receding and approaching sides.

As mentioned in the introduction, the main assumption about deriving rotation curves is that they are azimuthally symmetric, which results in a symmetric mass distribution. However, as our analysis shows, there can be huge deviations between the rotation curves extracted from the receding and approaching sides. As the occurrence of asymmetries in disc galaxies is quite high, extra care has to be taken when working on rotation curves and the mass distribution in those objects.
\section{Summary}
We obtained rotation curves of a sample of 70 spiral and irregular galaxies selected from the WHISP survey by performing a tilted-ring analysis of the \HI\ data. We looked at receding and approaching sides separately and discussed the differences in the velocities. The galaxies can be divided into five different types with respect to the behaviour of the rotation curves of receding and approaching sides and indicate symmetric velocity fields, global distortions or local distortions at different radii. The most common types are Type~2 where the velocities of receding and approaching sides show a constant offset over all radii, often as high as 50\skms, and Type~5 where the curves of receding and approaching sides change sides. Both types of galaxies are either members of a group of close companions, or have a close companion of higher or lower mass or an \HI\ cloud nearby. We suggest that these global distortions are the result of the disc response to a halo that was distorted by a tidal encounter.

Furthermore, we calculated the perturbation parameter for the lopsided potential from the kinematic asymmetry, $\epsilon_{\rm kin}$, which gave us a mean value of about 0.056 (averaged over all sample galaxies). However, this parameter either under- or overestimates lopsidedness if there are local distortions.

In order to quantify lopsidedness at all radii and to get a better idea of its physical origin(s), a more detailed study of lopsidedness has to be carried out. Morphological lopsidedness can easily be measured on all scales. A detailed study of the morphological lopsidedness will allow us to shed some light on the mechanisms that lead to lopsidedness in general. This analysis will be done in Paper II.
\begin{acknowledgements}
The authors would like to thank the anonymous referee for the constructive feedback which helped to improve this paper.\\
We want to thank Thijs van der Hulst for providing us with the WHISP data cubes before they became publicly available. C.~J. would like to thank DFG (Germany) and INSA (India) for supporting a visit to Germany in October 2007 under INSA-DFG Exchange Programme, during which this collaboration was started. We made extensive use of NASA's Astrophysics Data System (ADS) Bibliographic Services and the NASA/IPAC Extragalactic Database (NED) which is operated by the Jet Propulsion Laboratory, California Institute of Technology, under contract with the National Aeronautics and Space Administration. We acknowledge the usage of the HyperLeda database (http://leda.univ-lyon1.fr).
\end{acknowledgements}
\bibliographystyle{aa}
\bibliography{bibliography}
\newpage
\begin{appendix} 
\section{Rotation curves of all sample galaxies}
\begin{figure*}
 \centering
\includegraphics[width=.99\textwidth,viewport=58 93 540 723, clip=]{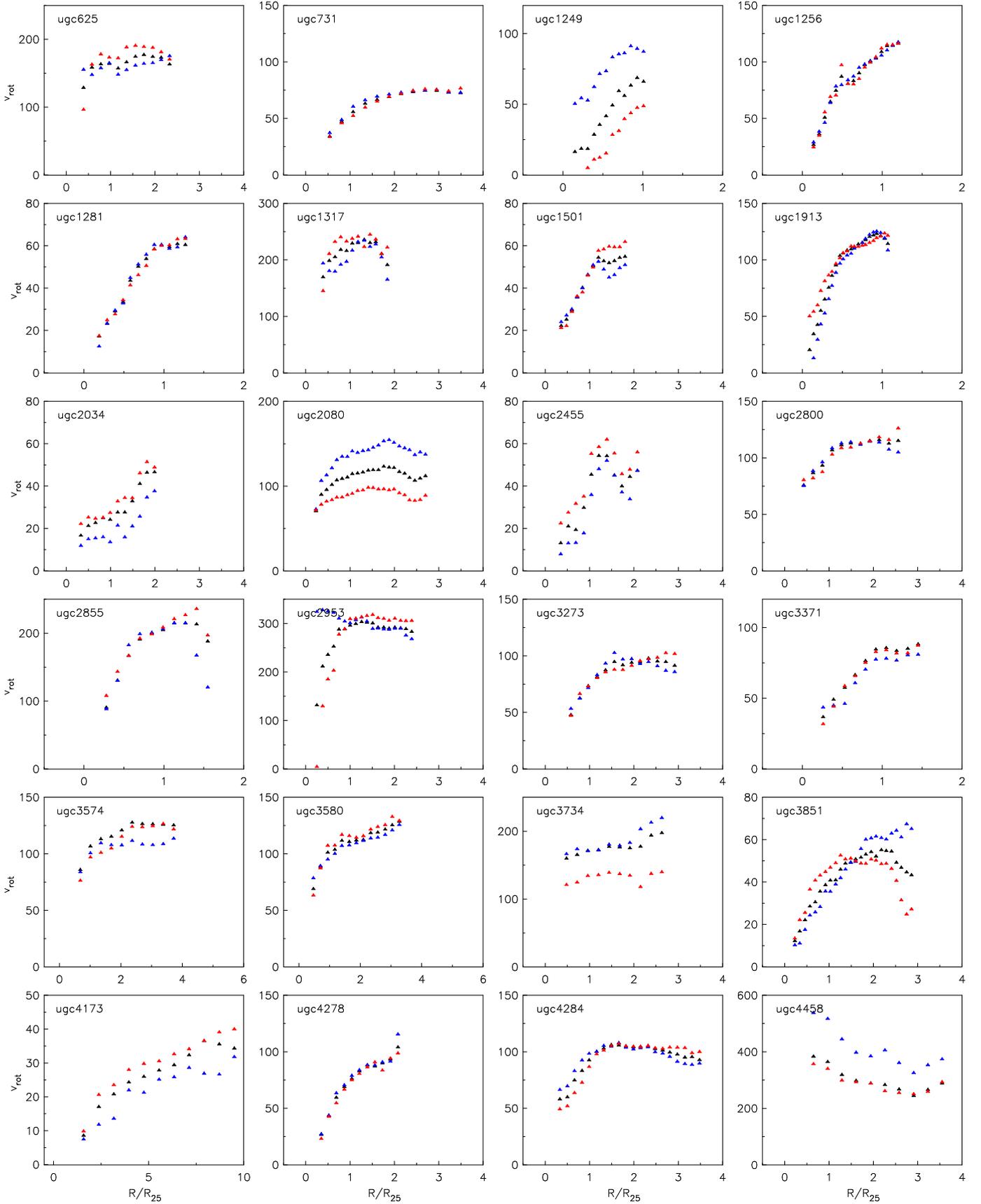}
\caption[]{The rotation curves of all sample galaxies extracted from the receding (red), approaching (blue), and a combination of both sides (black).}
\label{fig_rotcur1}
\end{figure*}

\begin{figure*}
 \centering
\includegraphics[width=.99\textwidth,viewport=58 93 540 723, clip=]{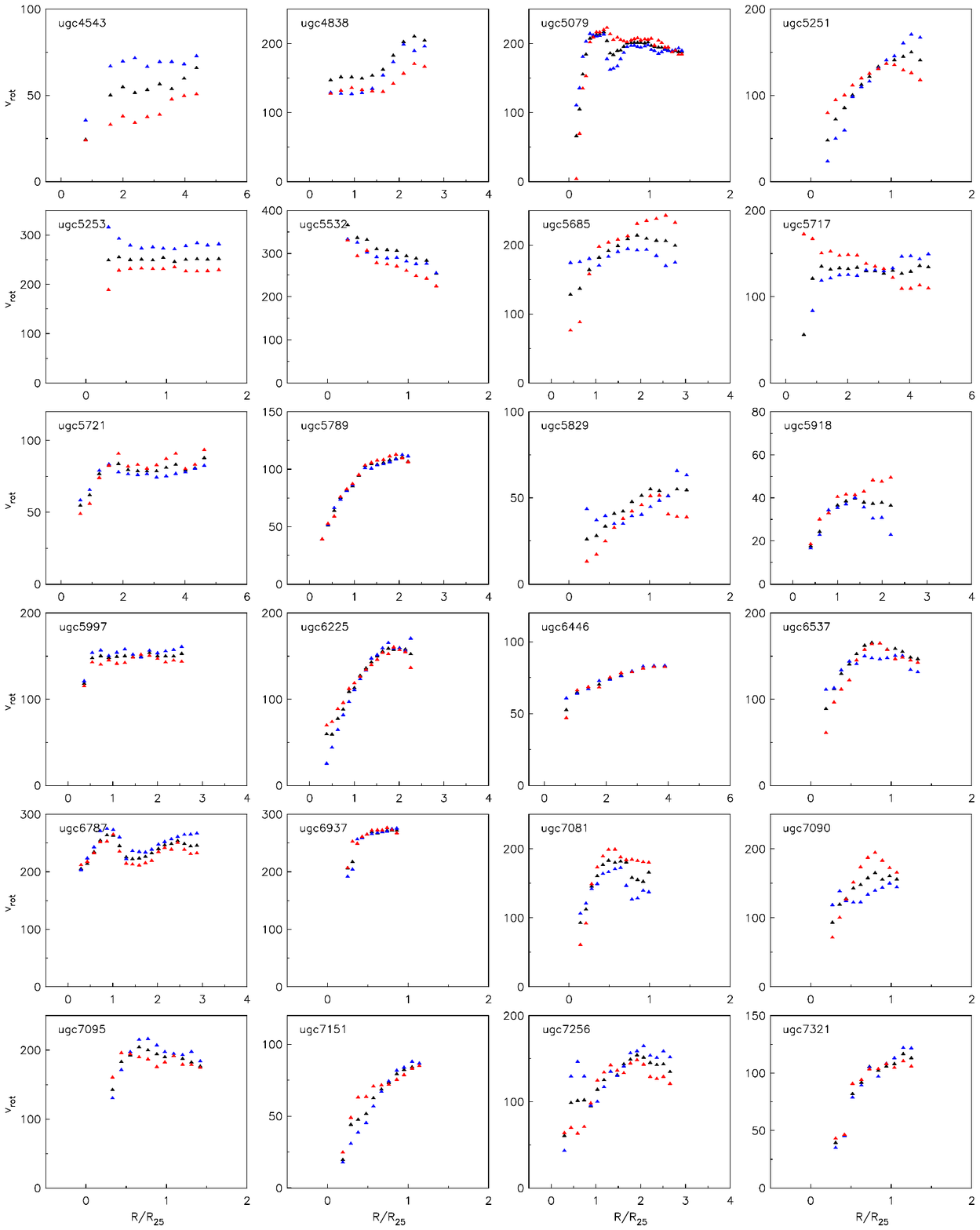}
\caption[]{Figure~\ref{fig_rotcur1} to be continued.}
\label{fig_rotcur2}
\end{figure*}

\begin{figure*}
 \centering
\includegraphics[width=.95\textwidth,viewport=58 93 540 723, clip=]{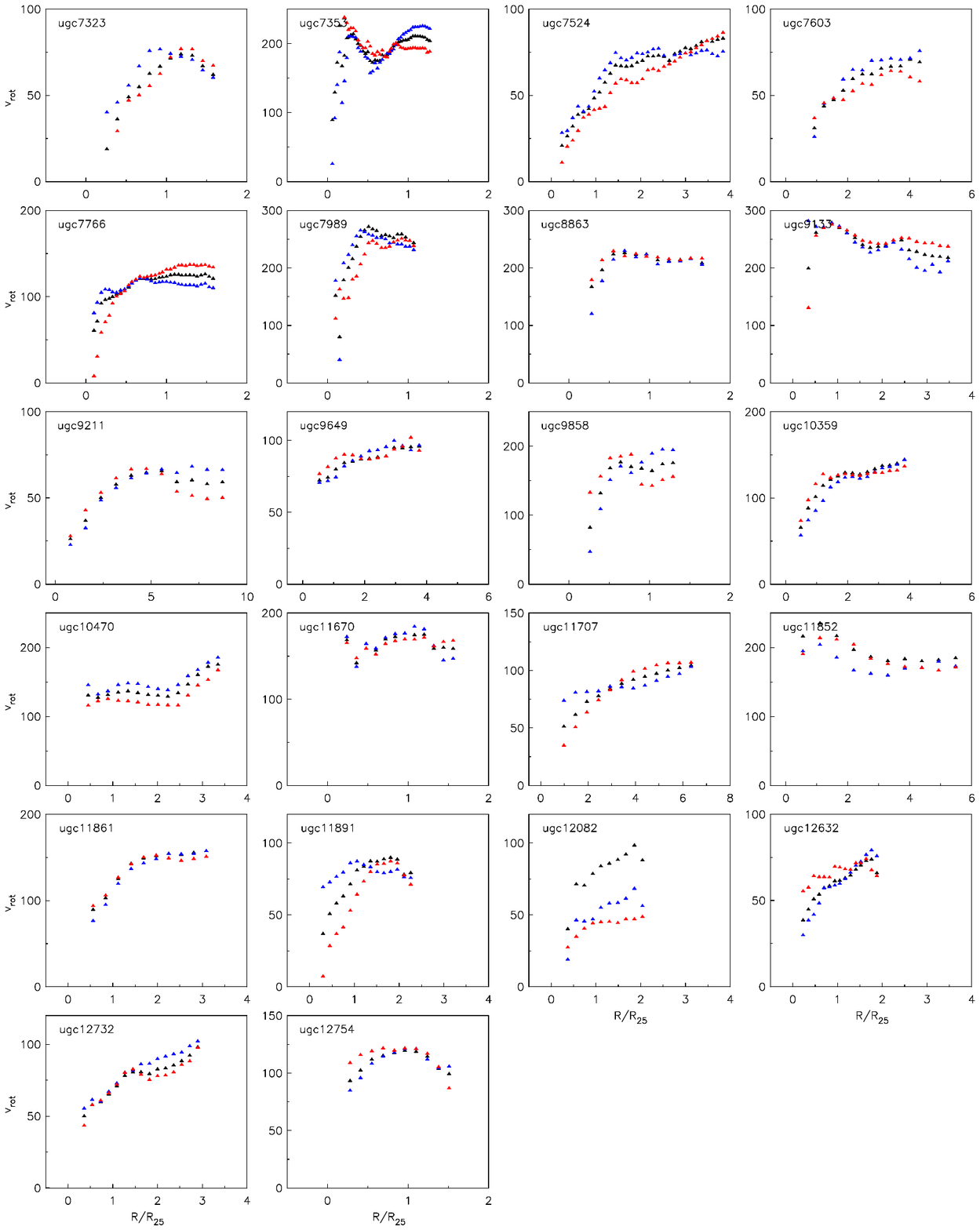}
\caption[]{Figure~\ref{fig_rotcur1} to be continued.}
\label{fig_rotcur3}
\end{figure*}

Here we present the rotation curves of all sample galaxies (Figs.~\ref{fig_rotcur1} to \ref{fig_rotcur3}). Red and blue symbols mark the rotation velocities of receding and approaching sides respectively. The rotation curve of a combination of receding and approaching sides is shown in black. 
\end{appendix}

\end{document}